# DenseLens – Using DenseNet ensembles and information criteria for finding and rank-ordering strong gravitational lenses

Bharath Chowdhary Nagam,[1]⋆ Léon V. E. Koopmans,[1] Edwin A. Valentijn,[1] Gijs Verdoes Kleijn,[1] Jelte T. A. de Jong,[1] Nicola Napolitano,[2,3,4] Rui Li[5,6] and Crescenzo Tortora[2]

[1]*Kapteyn Astronomical Institute, University of Groningen, PO Box 800, NL-9700 AV Groningen, the Netherlands*
[2]*INAF – Osservatorio Astronomico di Capodimonte, Via Moiariello 16, I-80131 Napoli, Italy*
[3]*School of Physics and Astronomy, Sun Yat-sen University, Zhuhai Campus, 2 Daxue Road, Xiangzhou District, Zhuhai 519082, China*
[4]*CSST Science Center for Guangdong-Hong Kong-Macau Great Bay Area, Zhuhai 519082, China*
[5]*School of Astronomy and Space Science, University of Chinese Academy of Sciences, Beijing 100049, China*
[6]*National Astronomical Observatories, Chinese Academy of Sciences, 20A Datun Road, Chaoyang District, Beijing 100012, China*

**ABSTRACT**
Convolutional neural networks (CNNs) are the state-of-the-art technique for identifying strong gravitational lenses. Although they are highly successful in recovering genuine lens systems with a high true-positive rate, the unbalanced nature of the data set (lens systems are rare), still leads to a high false positive rate. For these techniques to be successful in upcoming surveys (e.g. with *Euclid*) most emphasis should be set on reducing false positives, rather than on reducing false negatives. In this paper, we introduce densely connected neural networks (DenseNets) as the CNN architecture in a new pipeline-ensemble model containing an ensemble of classification CNNs and regression CNNs to classify and rank-order lenses, respectively. We show that DenseNets achieve comparable true positive rates but considerably lower false positive rates (when compared to residual networks; ResNets). Thus, we recommend DenseNets for future missions involving large data sets, such as *Euclid*, where low false positive rates play a key role in the automated follow-up and analysis of large numbers of strong gravitational lens candidates when human vetting is no longer feasible.

## 1 INTRODUCTION

Strong gravitational lensing is a phenomenon by which a massive foreground object (acting as a lens) distorts the light path from a more distant source into distinct (resolved) multiple images (Kochanek 2006; Treu 2010; Congdon & Keeton 2018). The geometry of lensed images may be multiple images of the source, an arc, or a ring depending on the nature of the source and alignment. Time delays between multiple images can be used to estimate the Hubble constant $H_0$ (Rhee 1991; Kochanek 2003; Grillo et al. 2018). Strong lensing also (i) acts as a high-resolution telescope without which many of the sources such as relativistic jets and supermassive black holes cannot be resolved by direct observations (Richard et al. 2014; Barnacka 2018), (ii) provides constraints on the dark energy density in the Universe (Sereno 2002; Linder 2004; Meneghetti et al. 2005; Biesiada 2006; Sarbu, Rusin & Ma 2001), (iii) is used to study the mass distribution of galaxies (Halkola, Seitz & Pannella 2006; Verdugo, de Diego & Limousin 2007; Nightingale et al. 2019) and dark matter (Treu & Koopmans 2004; Barnabè et al. 2009), and (iv) provides constraints on the slope of inner mass density profile (for e.g. Treu & Koopmans 2002; Gavazzi et al. 2007; Koopmans et al. 2009; Zitrin et al. 2012; Spiniello et al. 2015; Li, Shu & Wang 2018).

Many strong lensing surveys such as the Cosmic Lens All-Sky Survey (CLASS) based on radio imaging (Browne et al. 2003), SDSS Quasar Lens Search (SQLS; Oguri et al. 2006, 2008) based on the spectroscopy method, the COSMOS survey (Faure et al. 2008; Jackson 2008) using *HST* images etc., have been conducted with each survey yielding a few to a few dozen lenses. The Sloan Lens ACS (SLACS) survey and Bolton et al. (2006) targeted early-type lens galaxies that had faint lensed sources and found 70 highly probable strong lensing candidates (Bolton et al. 2008). Shu, Bolton & Brownstein (2015) and Shu et al. (2017) further extended the list of lensing candidates found with SLACS to around 100. Strong lenses have also been found at sub-mm wavelengths with the South Pole Telescope (SPT; Bleem et al. 2015) and at near-infrared wavelengths (McKean et al. 2007). Another more than a thousand strong lensing candidates have been found (Chan 2016; More et al. 2016; Nord et al. 2016; Tanaka 2016; Diehl et al. 2017; Treu et al. 2018; Jacobs et al. 2019a, b; Rojas et al. 2022) with ground-based surveys such as the Dark Energy Survey (DES; The Dark Energy Survey Collaboration 2005), the Canada–France–Hawaii Telescope Lensing Survey (CFHTLens; Heymans et al. 2012), the Hyper Suprime-Cam Survey (Miyazaki et al. 2012), the Kilo-Degree Survey (KiDS; de Jong et al. 2013) by Petrillo et al. (2017, 2019b), Li et al. (2020, 2021), and the VST Optical Imaging of the CDFS and ES1 fields (VOICE; Gentile et al. 2021).

Upcoming large sky surveys, for example with the Vera C. Rubin Observatory (previously referred to as the Large Synoptic Survey

⋆ E-mail: b.c.nagam@rug.nl

Telescope, *LSST*; Tyson 2002), *Euclid* (Laureijs et al. 2010), the Square Kilometer Array (SKA; Dewdney et al. 2009; Quinn et al. 2015), and the *Chinese Space Station Telescope* (*CSST*; Zhan 2018) are expected to discover another $10^5$ strong lenses (e.g. Pawase et al. 2014; Serjeant 2014; Collett 2015) each.

Traditional techniques to find lenses such as visual inspection or other previously applied algorithms will not optimally work on these large data sets, due to the size of the data set and the diversity of lens systems. Due to recent success in classifying lenses using convolutional neural network (hereafter CNN) in the strong gravitational lens finding challenge (Metcalf et al. 2019), they have become the preferred search technique, being both fast and flexible.

A CNN is a gradient-based learning algorithm. It was first introduced in 1998 by LeCun (Lecun et al. 1998) for handwritten digit recognition. Later, CNNs outperformed all other models in the ImageNet Large Scale Visual Recognition Challenge (ILSVRC). For example, AlexNet (Krizhevsky, Sutskever & Hinton 2012), won the 2012 ILSVRC challenge by achieving a top-5 error rate of 15.3 per cent in classifying the ImageNet data set. It uses data augmentation and dropout as a form of regularization techniques. In 2014, GoogLeNet (Szegedy et al. 2014) won the ILSVRC challenge, pushing the error rate below 7 per cent. Residual Network (ResNets; He et al. 2016) won the 2015 ILSVRC challenge, with an error rate under 3.6 per cent.

Due to its success, CNNs have been applied to find strong lenses by Petrillo et al. (2017, 2019a, b), Pearson, Pennock & Robinson (2018), Davies, Serjeant & Bromley (2019), Metcalf et al. (2019), Li et al. (2020, 2021), and Rezaei et al. (2022). In a recent strong gravitational lens finding challenge (Metcalf et al. 2019), different machine learning algorithms and deep learning algorithms (such as SVM, ResNets, AlexNets) have been used.

Although the above-mentioned techniques are extremely successful, they also possess considerable risk of overfitting the training data due to the relatively large number of parameters (often tens of millions) to be optimized during the training of the networks. DenseNets (Huang et al. 2017) address this problem by drastically reducing the number of parameters using 'dense connectivity patterns'. The dense connectivity pattern uses feature maps of all preceding convolutional layers that makes them parameter efficient.

Whereas in ResNets, features are combined through summation before passing on to a layer, in DenseNets, features are concatenated. The latter is the major difference between DenseNets and ResNets contributing to the improved network efficiency of DenseNets, eventually reducing overfitting on small data sets. DenseNets also allow improved flows of information and gradients throughout the network, making them easier to train.

In this paper, we introduce DenseNets for the first time to detect strong gravitational lenses and compare the performances of DenseNets to ResNets in classifying simulated mock lenses and non-lenses. In Section 2, we describe the methods to build mock data and to rank-order them. In Section 3, we explain the architecture of the CNN pipeline used to classify and rank-order mock lenses and introduce different metrics to assess the performance of CNN. Finally, in Section 4, we explain our results and in Section 5, we provide our main conclusion with discussion.

## 2 DATA SETS AND INFORMATION CONTENT

In this section, we discuss the data from the Kilo Degree Survey (KiDS; de Jong et al. 2013) being used to train and assess the network as well as various metrics used to rank-order lenses. KiDS is a wide-field optical imaging survey with an OmegaCAM camera (Kuijken et al. 2011) on the VLT-Survey Telescope (VST; Capaccioli & Schipani 2011) in Chile. KiDS has observed about 1350 square degrees (Kuijken et al. 2019) in four filters (*u*, *g*, *r*, and *i* bands). The *r*-band images have a point spread function (PSF) full width at half-maximum (FWHM) of <0.7 arcsec and an exposure time of 1800 s. For the purpose of this paper, we use data from the *r*-band KiDS DR4 data release (Kuijken et al. 2019) containing 1006 tiles of around 1 square degree each, from which we use the first 904 tiles processed by AstroWISE. These data are identical to the data used previously by Petrillo et al. (2019b) and Li et al. (2020).

### 2.1 Simulated lensed systems

We follow Petrillo et al. (2017) and generate mock lensed systems by combining simulated lensed sources with images of observed galaxies. Luminous red galaxies (LRGs; Eisenstein et al. 2001) from KiDS DR4 data are selected to train the networks. LRGs are massive galaxies, which are more likely to exhibit strong lensing features. Low-redshift ($z < 0.4$) LRGs are selected by clipping areas in ($r - i$) ($g - r$) colour diagrams based on the following criteria:

$$|c_{\text{perp}}| < 0.2,$$
$$r < 14 + c_{\text{par}}/0.3,$$

where (1)

$$c_{\text{par}} = 0.7(g - r) + 1.2[(r - i) - 0.18],$$
$$c_{\text{perp}} = (r - i) - (g - r)/4.0 - 0.18.$$

In total, 5514 LRGs were obtained which are split into 4411 training samples, 552 validation samples, and 551 test samples. The training samples are used to train the CNN by updating its parameters. Validation samples are not used for training the CNN, but to prevent the CNN from overfitting. In addition, we have also used augmentation techniques to prevent overfitting such as random rotation, flipping and rescaling identical to that by Petrillo et al. (2019a). The augmentation is applied to both sets of mock lenses and non-lenses during the training phase of the convolutional neural nets. The test samples are used to evaluate the CNN after the training is completed. We refer the interested reader to Petrillo et al. (2019a) for more detailed explanations on creating training samples.

### 2.2 Contaminants

There are many objects in the Universe that can mimic a strong lens system, given the limited depth and resolution of the imaging and colour information. It is critical that the CNNs are trained to better recognize these contaminants. Contaminants are selected KiDS images containing mergers, spirals, galaxies with dust lanes etc., which are used to train the CNNs as non-lenses. We have used the same data set for contaminants as discussed by Petrillo et al. (2019a). Samples of contaminants are shown in the second row of Fig. 1. The systems were selected as follows:

(i) ∼3000 galaxies having *r* magnitude <21 are randomly chosen to train the network with general true negatives.

(ii) ∼2000 sources, having been wrongly identified as mock lenses in previous tests (Petrillo et al. 2017).

(iii) ∼1000 galaxies have been visually classified spiral galaxies from GalaxyZoo project (Lintott et al. 2008, 2011; Willett et al. 2013; Melvin et al. 2014).

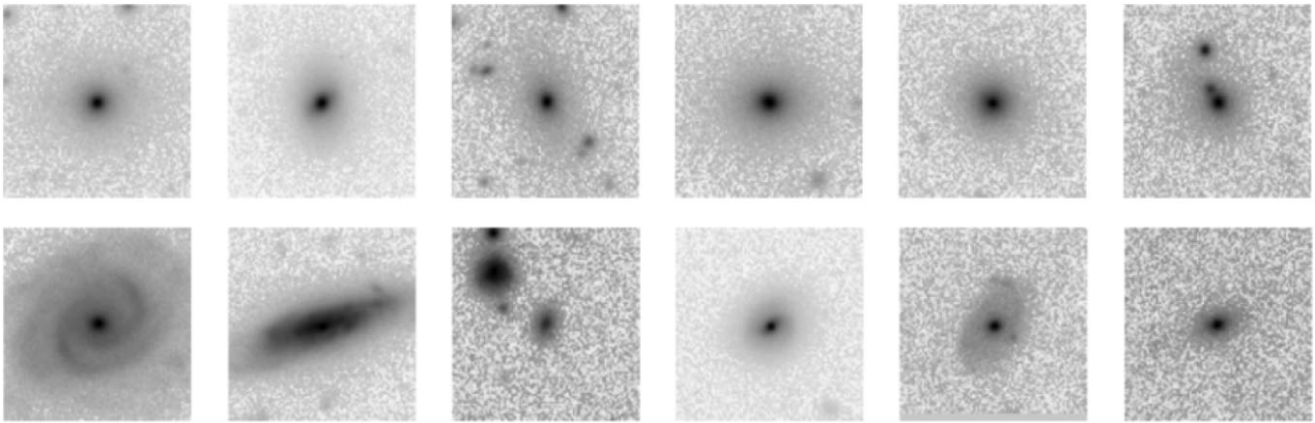

**Figure 1.** Examples of the non-lenses in the training data set. Figures in the first row are LRGs and in the second row are contaminants (negatives).

**Table 1.** Range of parameter values for simulating the lensed sources (Petrillo et al. 2019a).

| Parameter | Range | Unit |
|---|---|---|
| Lens (SIE) | | |
| Einstein radius | 1.0–5.0 | arcsec |
| Axial ratio | 0.3–1.0 | – |
| Major-axis angle | 0.0–180 | degree |
| External shear | 0.0–0.05 | – |
| External-shear angle | 0.0–180 | degree |
| Main source (Sérsic) | | |
| Effective radius ($R_{\rm eff}$) | 0.2–0.6 | arcsec |
| Axial ratio | 0.3–1.0 | – |
| Major-axis angle | 0.0–180 | degree |
| Sérsic index | 0.5–5.0 | – |
| Sérsic blobs (1 up to 5) | | |
| Effective radius | (1–10 per cent)$R_{\rm eff}$ | arcsec |
| Axial ratio | 1.0 | – |
| Major-axis angle | 0.0 | degree |
| Sérsic index | 0.5–5.0 | – |

During the training, the above ∼5000 sources are used as examples of false positives and classified as non-lenses. The remaining ∼1000 were split equally for validation and testing purposes. The augmentation techniques used for simulated lens systems are also applied to contaminants.

## 2.3 Mock lensed sources

Sources are modelled by sampling parameters from a Sérsic radial profile (Sérsic 1968) and the lenses with a Singular Isothermal Ellipsoid (SIE; Kormann, Schneider & Bartelmann 1994) as listed in Table 1. The parameter space listed in the table is the same as used by Petrillo et al. (2019a). Einstein radii of the lenses (in arcsec) and the effective (half total light) radii of the main sources have a logarithmic distribution while the other parameters have a flat distribution. In total, $10^5$ mock lensed sources of 101 × 101 pixels are simulated corresponding to a 20 arcsec × 20 arcsec area. To help the CNN to be able to recognize all possible lenses in the real Universe, we note that the distribution of parameters used in creating mock lenses is chosen to cover the potentially wide range of parameter space in reality. This affects the metrics of success of the networks as discussed later in the paper. Mock lenses are created by combining the simulated sources and the observed KiDS LRGs as shown in Fig. A1. We refer to Petrillo et al. (2017) for more details.

### 2.3.1 Information content and rank ordering of lens systems

We define the information content (IC) for each image to help CNN to rank-order them. The value of the IC scales linearly with the number of resolution elements (i.e. the area of the PSF) of the training noiseless mock lensed images above a given brightness threshold in units of the background noise ($\sigma$). This heuristic metric is motivated by the fact that each resolution element provides three pieces of information, two coordinates, and a brightness. We note that the information does not linear scale with brightness itself. One expects higher IC values to correspond to easier-to-recognize lens systems. The IC value is added as a metric to train the CNN algorithm and is also predicted again when a lens candidate is selected by the network above a certain threshold. Based on this IC, we can rank-order lenses to avoid human ranking. In practice, the IC of the simulated source image is defined as

$$\text{IC} = \left[\frac{A_{\rm src, 2\sigma}}{A_{\rm PSF}}\right] \times R, \quad (2)$$

In the above equation, $A_{\rm src, 2\sigma}$ is defined as the total area of the lensed images above a given brightness threshold in the unit of the background noise $\sigma$. We tried several thresholds and found the value of $2 \times \sigma$ to work well in practice. We define the area of PSF ($A_{\rm PSF}$) as the square of FWHM. $R = (R_{\rm E}/R_{\rm eff})$ is the ratio of the Einstein radius ($R_{\rm E}$) over the effective source radius ($R_{\rm eff}$). This extra factor in the IC helps to avoid candidates that have a large effective source radius and a small Einstein radius to have a very large IC value, despite having limited lensing features. We find that this extra correction factor helps in rank-ordering the lens candidates. The IC is not a rigorous definition in the context of information theory, but it is a metric of how easy we expect it is to recognize a lens system in the data for a human and for a neural network classifier.

## 3 THE CLASSIFICATION AND RANK-ORDERING NETWORKS

In this section, we describe the neural network that we use to classify and rank-order lens candidates. Rather than using a single neural network, we use so-called ensemble networks that each classify a system, and where a final classification is based on their joint result. Ensemble networks were first introduced by Rosen (1996) and it was shown that ensemble networks can have lower errors than individually trained networks. We use two types of networks in this paper.

In a classification network (CNNs 1–4), the output layer is made up of one dense neuron with Sigmoid activation (Narayan 1997) function which predicts values in the range of [0,1]. A threshold is set within this range where the candidates whose predictions fall above this threshold are classified as promising lens candidates and the others are classified as non-lenses.

In a regression network (CNNs 5–8), the output layer is made up of a linear activation function. The network is trained with IC values of training images. Mock lenses are trained with their respective IC values and non-lenses are trained with values equal to zero.

However, it is important to note that the training set is the same for both the classification and regression networks. The only difference is that the classification network (CNNs 1–4) uses binary values (0 or 1) for training and the regression networks (CNNs 5–8) use IC values for training as mentioned in Section 2.3.1.

CNNs are stochastic training algorithms and they often can differ in weights at the end of training, resulting in different predictions. Thus, the difference between each individual CNNs in CNNs 1–4 and CNNs 5–8 is that they have slight differences in weights even though they have been trained on the same data for the same duration of time and have the same architecture.

### 3.1 Network architecture

We use DenseNets (Huang et al. 2017) as the network architecture for both classification and regression networks. Huang et al. (2017) showed that DenseNets uses far fewer parameters to achieve the same level of accuracy as ResNets. Thus, in this paper, we compare the performance of the DenseNet-121 (∼1M params) ensemble with the performance of ResNet-18 (∼11.7M params) (He et al. 2016) ensemble networks and customized ResNet used in Li et al. 2021 (hereafter Li (2021) ResNet+). Li (2021) ResNet+ is a modified version of ResNet-18 and has two additional Dense layers of 512 units. This makes Li (2021) ResNet+ parameter heavy with ∼13M params. The DenseNet-121 architecture has 120 convolutional layers and 1 fully connected layer. DenseNet-121 network mentioned in Huang et al. (2017) has growth rate (k) equals 32 and ∼8M parameters. However, to improve feature sharing and to reduce the number of parameters, we have set growth rate (k) to 12 and we have used Bottleneck (B) layers and Compression (C) as suggested in Huang et al. (2017). Thus, the architecture we have used is a DenseNet-BC variation with 121 layers and it has only ∼1M parameters. Thus, an ensemble of four DenseNet-121 networks has ∼4M parameters in total. Interested readers can look into Appendices B and C for further information about DenseNets and ResNets.

### 3.2 DenseLens: Pipeline-Ensemble model

We use the classification and regression networks sequentially, where a system is first classified, and only when it exceeds a minimum threshold for being a promising lens candidate, its IC value is predicted from the data to rank-order it. However, during training, the regression networks were trained independently before being used in the Pipeline-Ensemble model. Our Pipeline-Ensemble model makes use of an ensemble of classification and regression networks arranged sequentially, as shown in Fig. 2. We call this Pipeline-Ensemble model the *DenseLens*, DenseNets for finding and rank-ordering strong gravitational lenses. We use an ensemble of four networks to reduce overfitting. Increasing the number of networks in the ensemble model increases the computational requirements and also increases the latency time to make each prediction. Hence, we have chosen the number of networks in an ensemble network to be four, balancing increased computational effort against improved classification. Images enter the classification ensemble network consisting of four trained classification CNNs (CNNs 1–4) which predict output values (P) in the range [0,1]. The mean of predictions ($P_{mean}$) is then computed, which are further discussed in detail in Section 4.1. Images having $P_{mean} > P_{thres}$ are selected as lens candidates. CNNs 5–8 are trained with IC values for lenses and zeros for all non-lenses. Each CNN in the network predicts a real number in the range [0, max(IC)]. The mean of the predicted IC values are calculated as $P_{IC\,mean}$ and the images are rank-ordered based on the value of $P_{IC\,mean}$.

We have tried other approaches such as combined CNN and concatenated CNN (as discussed in Appendix F) for classifying mock lenses. We tried using one CNN with two outputs (classification and regression) and two independent loss functions in the combined model. We have also tried a combined loss function for two outputs in the concatenated model. We have also tried a Cascade-type classifier for classification. We have found that the other approaches were not as efficient as the Pipeline-Ensemble model.

### 3.3 Training and testing the network

We use 'DenseNet-121' model (as detailed in Appendix B) to classify lens candidates. DenseNets have been recently used in the studies of strong gravitational lensing to probe the substructure of dark matter haloes (Alexander et al. 2020) and to identify strong lensing gravitational wave events (Goyal et al. 2021). We introduce DenseNets here to detect strong gravitational lenses. We define our problem as a binary classification problem. Hence, the CNNs in the classification network need to be trained on positive and negative samples. Positive samples are created by combining the LRGs with mock lensed sources. The procedure follows Petrillo et al. 2017.

An LRG (Section 2.1) and a mock lensed source (Section 2.3) are randomly selected. The peak brightness of the LRG in the *r* band is multiplied by a factor $\alpha$ randomly drawn from the interval [0.02,0.3] to set the peak brightness of the source. This enables one to lower the brightness of lensed-source features with respect to the LRG galaxy. An example of a mock lens is shown in the Fig. A1. To enhance lower luminosity features, negative-value pixels in all images are set to zero and a square-root stretch of the image is performed. The resulting image is normalized by the galaxy peak brightness.

To create non-lenses (see also Petrillo et al. 2017), negative samples are created by randomly choosing a galaxy from an LRG sample (with a 20 per cent probability) or by randomly choosing one of the earlier-selected false positives (with an 80 per cent probability). Also here, negative values pixels are clipped to zero and a square-root stretch of the image is performed. The resulting image is normalized by the galaxy peak brightness. The test set contains 10 000 mock lenses and 10 000 non-lenses.

To train the regression CNN networks, the IC values are determined for each of the simulated strong lensing images. The regression CNN networks (CNNs 5–8 Fig. 2) are trained with the IC values of the simulated strong lenses. The non-lenses are trained with values equal to zero. The model of the trained regression network is then used to predict the IC values of images. Based on these predicted IC values, the images are rank-ordered, avoiding any human inspection and rank-ordering. The results of this will be discussed in Section 4.

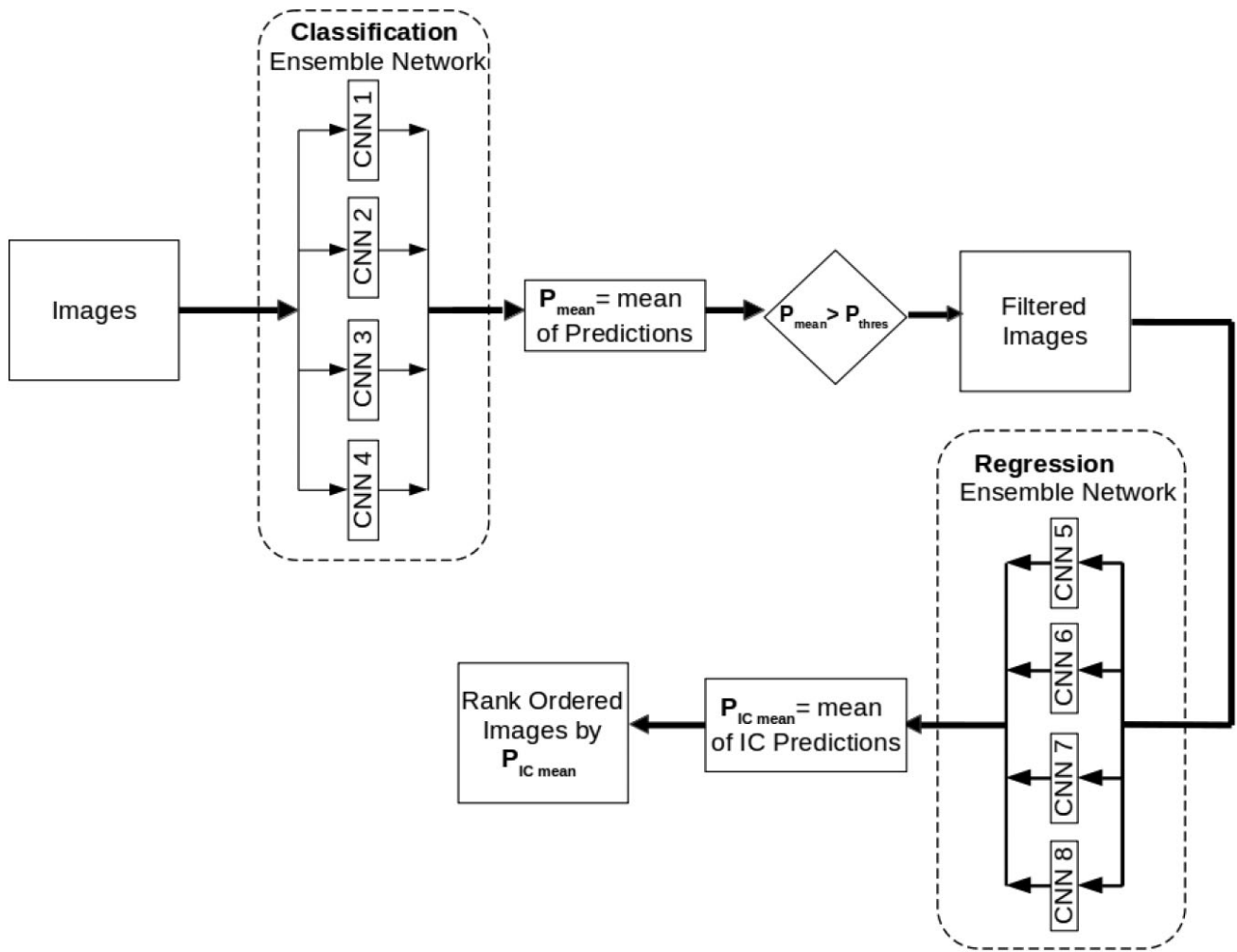

**Figure 2.** Description of Pipeline-Ensemble model. Images are selected by the four classification CNN networks (CNN1, CNN2, CNN3, and CNN4) which are above the threshold ($P_{thres}$). Selected images are passed on to four regression networks (CNN5, CNN6, CNN7, and CNN8). Each CNN in the regression networks predicts the value of IC ($P_{IC\,mean}$). Mean value of these IC predictions are used to rank-order the images.

We have also used the same training, validation samples to train and validate the ResNet-18 ensemble networks and Li (2021) ResNet+ network.

## 4 RESULTS

Having trained the network in Section 3, we explain the prediction of the classification network in Section 4.1. We also explain the receiver operator characteristic (ROC) curve in Section 4.2 and rank-ordering of lenses in Section 4.3. Finally, in Section 4.4 we explain the impact of our architecture on the upcoming *Euclid* mission.

### 4.1 Distribution of classification prediction

The distribution of classification predictions ($P_{mean}$) by DenseNet and ResNet-18 are shown for both lenses and non-lenses in Fig. 3 (top). In an ideal scenario, the values of $P_{mean}$ should be equal to 1 for lenses and 0 for non-lenses. Although DenseNet-121 ensemble classifies more lenses in the range between 0 and 0.3 than ResNet-18 ensemble, DenseNet-121 ensemble shows a sharper decline in the number of non-lenses at high $P_{mean}$ values when compared to ResNet-18 ensemble. Hence, DenseNet has far fewer false positives above the selection threshold for lenses set typically at $P_{mean} > 0.5$ (but often much closer to 1.0).

The distribution of classification predictions ($P_{mean}$) by DenseNet-121 ensemble and Li (2021) ResNet+ are shown for both lenses and non-lenses in Fig. 3 (bottom). Here, we can observe that by setting the $P_{mean}$ threshold to 0.95, DenseNet-121 ensemble selects slightly fewer lens systems than Li (2021) ResNet+, but effectively reduces the number of non-lenses by a factor of $\sim 7$ per cent. In highly imbalanced data sets, where non-lenses outnumber lenses by several orders of magnitude, this can lead to a drastic reduction of false positives. The latter drives current improvements in automated lens selections.

### 4.2 ROC curve

The performance of the networks can be further assessed using two different metrics, namely the true positive rate (TPR) and the false positive rate (FPR; Jones & Athanasiou 2005). The TPR is defined as

$$\text{TPR} = \frac{\text{TP}}{\text{TP} + \text{FN}} \quad \epsilon \ [0, 1], \tag{3}$$

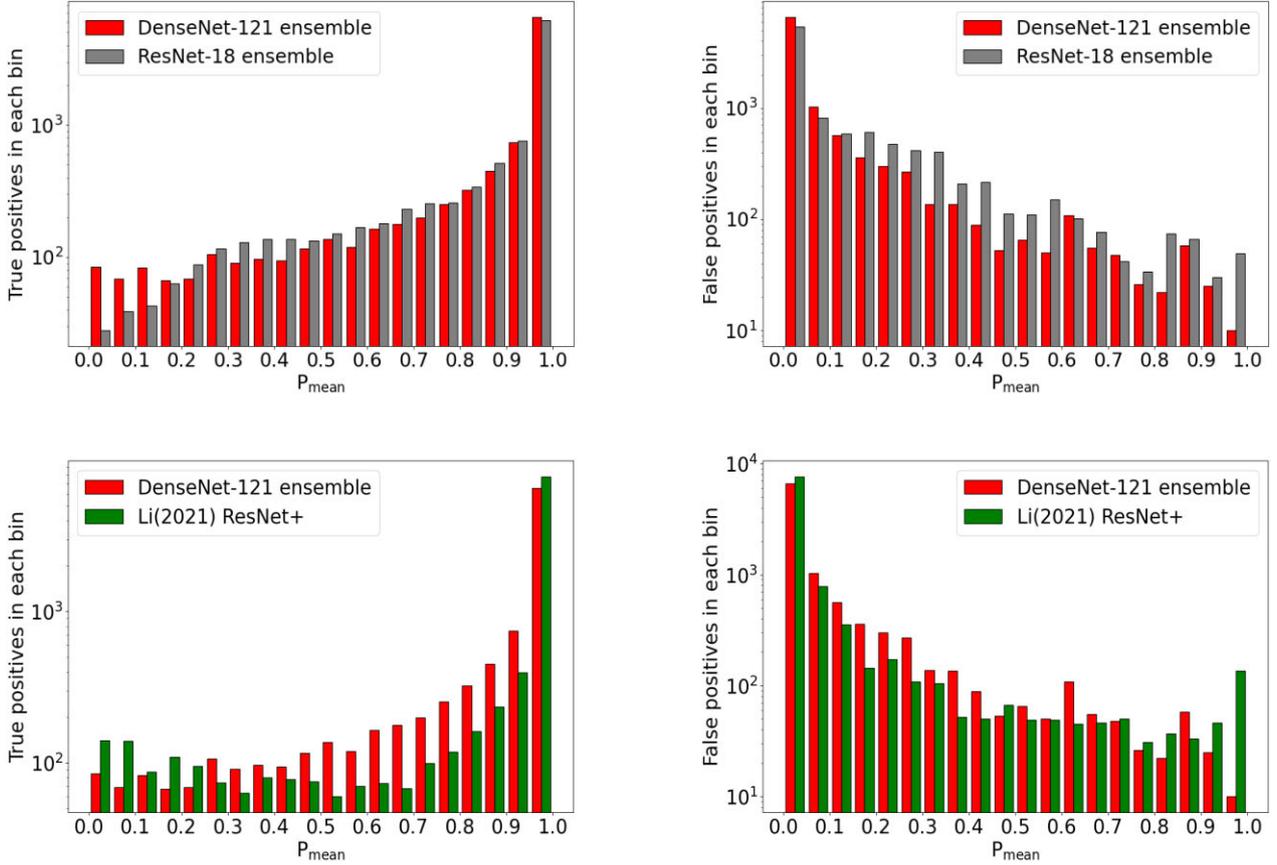

**Figure 3.** The distribution of classification prediction ($P_{mean}$) values of *mock lenses* (top left) and *non lenses* (top right) assigned by DenseNet-121 ensemble and ResNet-18 ensemble architectures for the same test sample. Similarly, the distribution of $P_{mean}$ values of *mock lenses* (bottom left) and *non-lenses* (bottom right) assigned by DenseNet-121 ensemble and Li (2021) ResNet+ architectures for the same test sample. The bottom left and right plot show the distribution in terms of true and false positives, respectively. The *y*-axis is log scaled in both the figures.

and is the ratio between the number of true positives (i.e. genuine lens systems) and the sum of true positives and false negatives (lens systems that the algorithm does not correctly identify). The FPR is defined as

$$\text{FPR} = \frac{\text{FP}}{\text{TN} + \text{FP}} \quad \epsilon \quad [0, 1], \quad (4)$$

and is the ratio between false positives (non-lenses falsely identified as genuine lenses) and the sum of true negatives (non-lenses) and false positives. These metrics are significant in building a ROC curve, which in turn are useful in visualizing the performance of our classification network.

To further analyse the performance of the two networks, TPR and FPR curves are used to build so-called ROC curves. Any point to the left of the diagonal line of a ROC plot will have a TPR greater than the FPR. $P_{mean}$ values can be tuned based on this curve to get an acceptable level of false positives and false negatives. The area under the curve (AUC) determines how efficient a model is compared to the rest of the other models. When the AUC value is higher, the performance of the model is better at classifying the positive and negative classes correctly. Thus, the AUC is often used as a metric to compare two different models. We note here, however, that in highly unbalanced data sets, in general, the ROC needs to cross a diagonal line not with a slope of 1, but a slope that reflects the imbalance. For example, if the number of non-lenses outweighs the number of lenses by a thousand to one, the ROC curve needs to rise extremely fast (high true positive rate for a very small false positive rate) in order not to contaminate the lens sample by too many non-lenses. Hence, besides the AUC, also the gradient of the ROC plays an important and often underestimated role.

Fig. 4 presents the ROC curves of ResNet-18 ensemble network, DenseNet-121 ensemble network and Li (2021) ResNet+ for the test data set described in Section 3.3. The ROC curve enables us to select an optimum threshold, i.e. the threshold where we can recover a high fraction of lenses without compromising too much on the false positive rate.

From Fig. 4, one can argue that 0.90, 0.94, or 0.97 are viable selection threshold values. On choosing 0.94 as an optimum threshold value over 0.90, we gain only 0.3 per cent in false positive rate and we lose approximately 6 per cent of true positives. This might initially seem to be counterintuitive. However, in reality, strong lenses are extremely rare and do not have a one-to-one ratio between lenses and non-lenses. For example, out of 100 000 sources typically one is a strong lens system. So a loss in 0.3 per cent false positives will imply that our final sample will be swamped by 300 more false positives for every strong lens found. Thus, we conclude that setting $P_{thres}$ at 0.94 is more optimal than setting $P_{thres}$ at 0.90. Similarly, we lose 7 per cent of true positives and we gain only a few false positives (only 0.04 per cent) by choosing 0.97 over 0.94 as an optimum threshold value. Hence, we have chosen 0.94 as our optimum threshold value ($P_{thres}$).

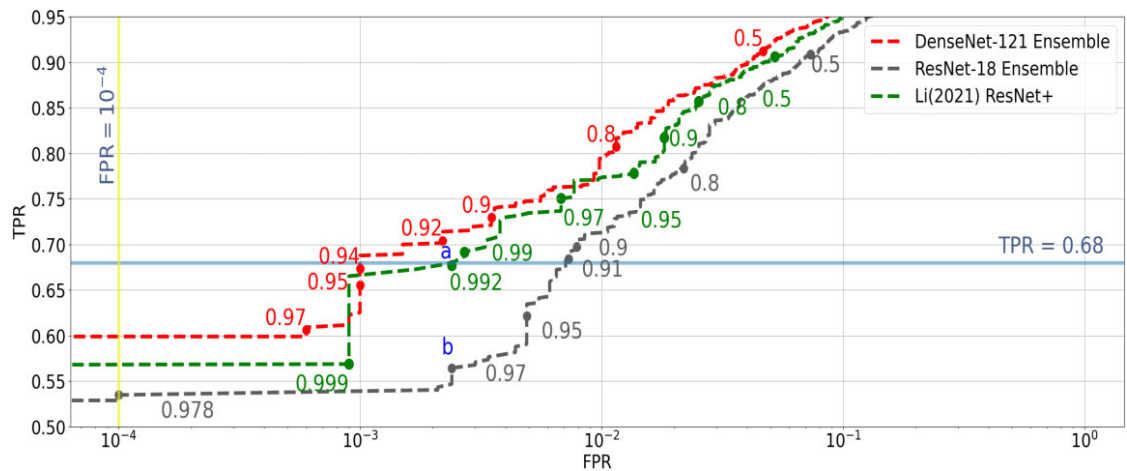

**Figure 4.** ROC curve of ResNet-18 ensemble, DenseNet-121 ensemble, and Li (2021) ResNet+ architectures. $P_{thres}$ values for respective ROC curves are shown as numerical text. $P_{thres} = 0.92$ for DenseNet-121 ensemble and $P_{thres} = 0.97$ for ResNet-18 ensemble are labelled as 'a' and 'b', respectively, for comparing the performance of two networks at similar FPR (explained in Section 4.4). When one sample out of 1000 samples is a mock lens (in a typical unbalanced data set), the number of mock lenses found can be compared one to one with non-lenses at a FPR rate of $10^{-3}$. At this low FPR of $10^{-3}$, the highest TPR achieved among all networks is 0.68 (achieved by DenseNet-121 ensemble). We compare the FPRs of three networks at constant TPR equal to 0.68 line that is shown as light blue horizontal line. Similarly, for upcoming large survey missions such as *Euclid*, it is important to compare network performances at extremely low false positive rates. Hence, we compare the TPRs of three networks at constant FPR = $10^{-4}$, which is shown as vertical yellow line.

In Fig. 4, we can compare the performance of the three networks at constant FPR and TPR. In an unbalanced data set, where typically one sample out of 1000 samples is a mock lens, the number of mock lenses found can be compared one to one with non-lenses at a FPR rate of $10^{-3}$. At this low FPR of $10^{-3}$, the highest TPR achieved among all networks is 0.68 (achieved by DenseNet-121). Thus, we can compare the FPRs of all three networks at the constant TPR of 0.68. DenseNet-121 ensemble, ResNet-18 ensemble, and Li (2021) ResNet+ achieve the constant TPR of 0.68 (shown as a light blue horizontal line) at $P_{thres}$ values of 0.94, 0.91, and 0.992 respectively. At these respective $P_{thres}$ values, the FPR of DenseNet-121 ensemble, ResNet-18 ensemble and Li (2021) ResNet+ are 0.001, 0.007, and 0.002 respectively. In other words, the DenseNet-121 ensemble network can recover 68 per cent of the mock lens with 2 times and 7 times fewer false positives when compared to the Li (2021) ResNet+ network and the ResNet-18 ensemble network. For upcoming large survey missions such as *Euclid*, having extremely low false positive rates are very important. DenseNet-121 ensemble, ResNet-18 ensemble, and the Li (2021) ResNet+ achieve very low FPR of $10^{-4}$ (shown as a yellow vertical line) at $P_{thres}$ values of 0.97, 0.978, and 0.999, respectively, and their respective true positive rates are 0.60, 0.54, and 0.57. So, at a very low false positive rate of $10^{-4}$, the DenseNet-121 ensemble recovers 3 per cent more mock lenses than Li (2021) ResNet+ and 6 per cent more mock lenses than the ResNet-18 ensemble network. In general, for any TPR considered, the DenseNet-121 ensemble has a smaller FPR when compared with the ResNet-18 ensemble and Li (2021) ResNet+. Also, we can observe that the ROC curves intersect only momentarily and the area under the curve (AUROC) is greater for DenseNet-121 ensemble network when compared with Li (2021) ResNet+.

The Li (2021) ResNet+ achieves better results when compared with ResNet-18 ensemble network (also with each individual ResNet-18 network as shown in Fig. E1 in the appendix). This is because the Li (2021) ResNet+ uses an additional two layers of 512 dense neurons in the end. This adds another half a million parameters to Li (2021) ResNet+ and improves its performance over ResNet-18 architecture. We can also observe that the DenseNet-121 ensemble network achieves this performance with 10 times fewer parameters when compared with ResNet-18 ensemble network or three times fewer parameters when compared with Li (2021) ResNet+.

Having shown that DenseNets achieves comparable performance with fewer parameters and a lower FPR value for fixed TPR value, we now turn our attention to automatically rank-ordering the lenses based on the predicted IC-values as defined in Section 4.3.

### 4.3 Rank ordering of lens candidates

Rank-ordered lens candidates are obtained as the output of our combined selection and rank-ordering network. Input images enter the network shown in Fig. 2. The images that have $P_{thres} > 0.94$ are selected by the values of the selected images and the IC values are predicted by the regression ensemble network ($P_{ICmean}$). The candidates are rank-ordered based on the IC value. The candidates having higher IC values (larger than 100) are, in general, correctly classified by the classification ensemble network. Fig. 5 (left) shows the relation between true IC (T) versus estimated IC (E) values for all the candidates in the test data set. The figure shows a strong correlation between the two parameters with root mean squared errors of about ∼100 and the Pearson correlation coefficient to be ∼0.92. Although the slope is not exactly one, we attribute this to a loss of information when the lensed images (based on which the true IC values are calculated) are injected into noisy data on top of a lens galaxy. Fig. 5 (right) describes the relation of E/T versus T. Here, we could observe that the scatter in E/T reduces as T increases. From this, we can conclude that as the information content increases, the correlation between estimated IC (E) and True IC (T) increases resulting in a decrease in scatter. The lens candidates are colour-coded by the classification prediction ($P_{mean}$) values as shown in the colour bar. From this colour coding, it is also clear that systems with a higher IC value (typically lenses with a larger Einstein radius) lead to larger *P*-values (i.e. they are easier to identify as lenses), as expected. This dependence on the network's ability to recover larger lens systems more easily will be further investigated later in the paper where we study the recovery rate as a function of Einstein radius.

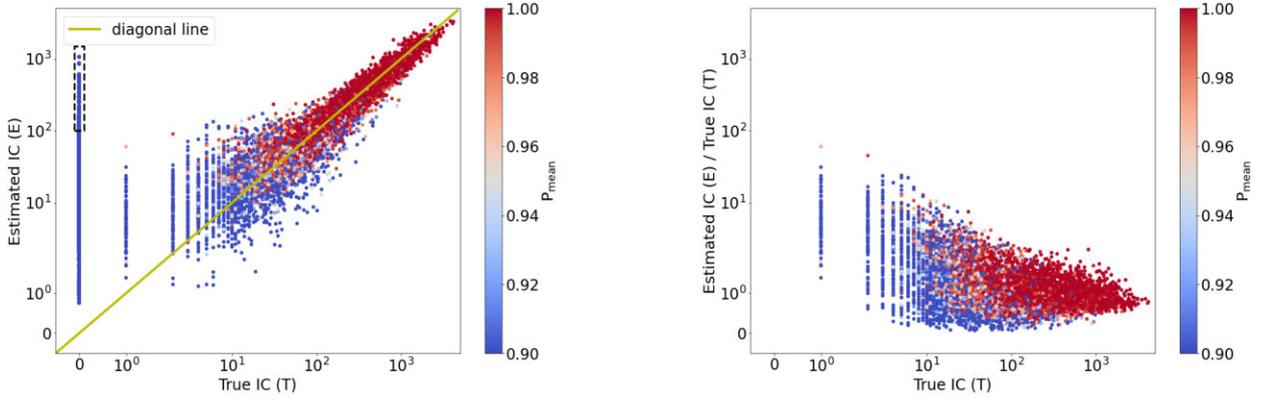

**Figure 5.** *Left:* Illustration of estimated IC (E) versus true IC (T) for all candidates in the test data set. The diagonal line is shown as a yellow line to guide the eye. *Right:* E/T versus T is plotted in the right figure. The $P_{mean}$ values are shown as colour bar in both right and left figures.

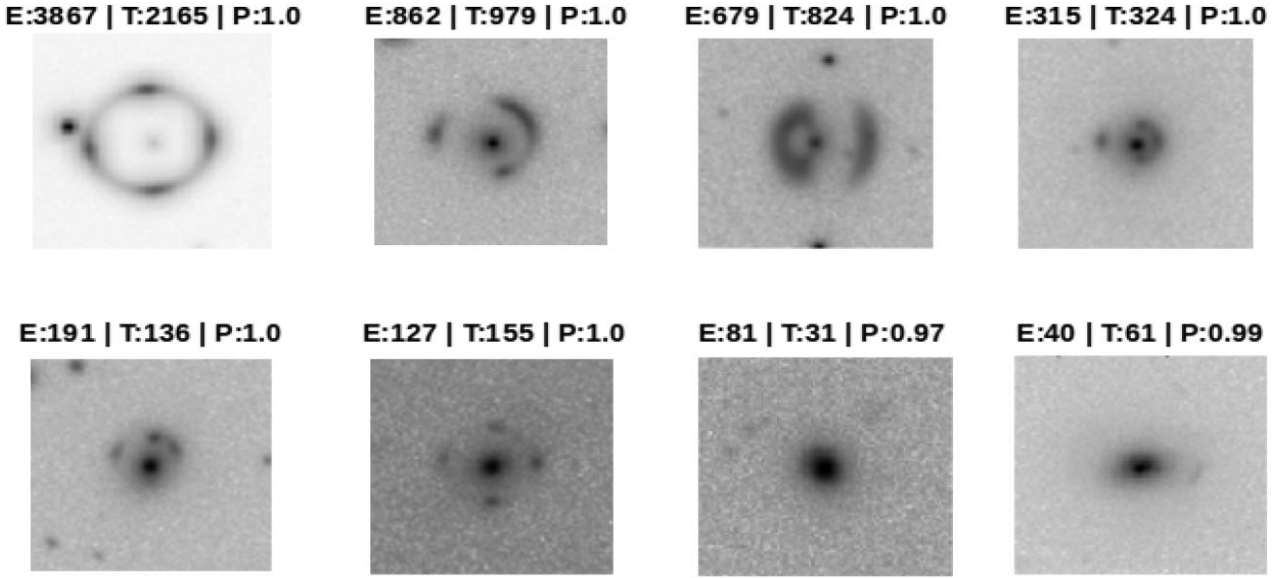

**Figure 6.** Shown are eight examples of rank-ordered images based on their estimated IC values (E). For each candidate, their true IC values (T) and the classification prediction scores (P) are also shown. We note that as the IC value decreases, in general, the Einstein radius decreases and the system becomes harder to identify visually.

In addition, it is clear that IC values larger than several tens are required for P to exceed 0.94. Given the resolution of KiDS data, used in the simulations, of about 0.7 arcsec, these must be systems with Einstein radii that are considerable (at least larger than 1 arcsec).

*4.3.1 Why a classification ensemble network is required?*

One can wonder why the regression ensemble network alone is sufficient and what is the need for a classification ensemble network at the beginning of the pipeline. There are a lot of candidates which are having true IC values equal to zero and their estimated IC values are very large. These candidates are shown inside a black dotted line rectangle and are false positives. When a classification ensemble network is used, it removes all the candidates less than or equal to 0.94 (shown as bluish dots) and the regression ensemble network will not rank-order these candidates. Thus, the final output will not be overwhelmed by the candidates having low true IC

(T) and large estimated (E) values. Each CNN network shows a strong correlation with the other CNN network in the classification ensemble networks (CNNs 1–4). Their maximum and minimum Pearson correlation coefficient were 0.88 and 0.84, respectively, for positive samples and 0.83 and 0.74, respectively, for negative samples. Similarly, regression ensemble networks (CNNs 5–8) also showed a strong correlation. The maximum and minimum Pearson correlation coefficient between individual network in CNNs 5–8 were 0.99 and 0.98, respectively, for positive samples and 0.93 and 0.85, respectively, for negative samples.

A panel with rank-ordered images, based on their predicted IC values, is shown in the Fig. 6. Each image is represented by its estimated IC values (E), its true IC values (T), and classification prediction (P) in its title. Estimated IC values (E) imply the prediction of regression CNN networks. True IC values (T) represent the true IC values calculated while generating the mock lenses and the classification prediction (P) represents the prediction of the DenseNet classification ensemble network. Rank-ordering of lens candidates is performed by ordering candidates based on estimated IC values (E).

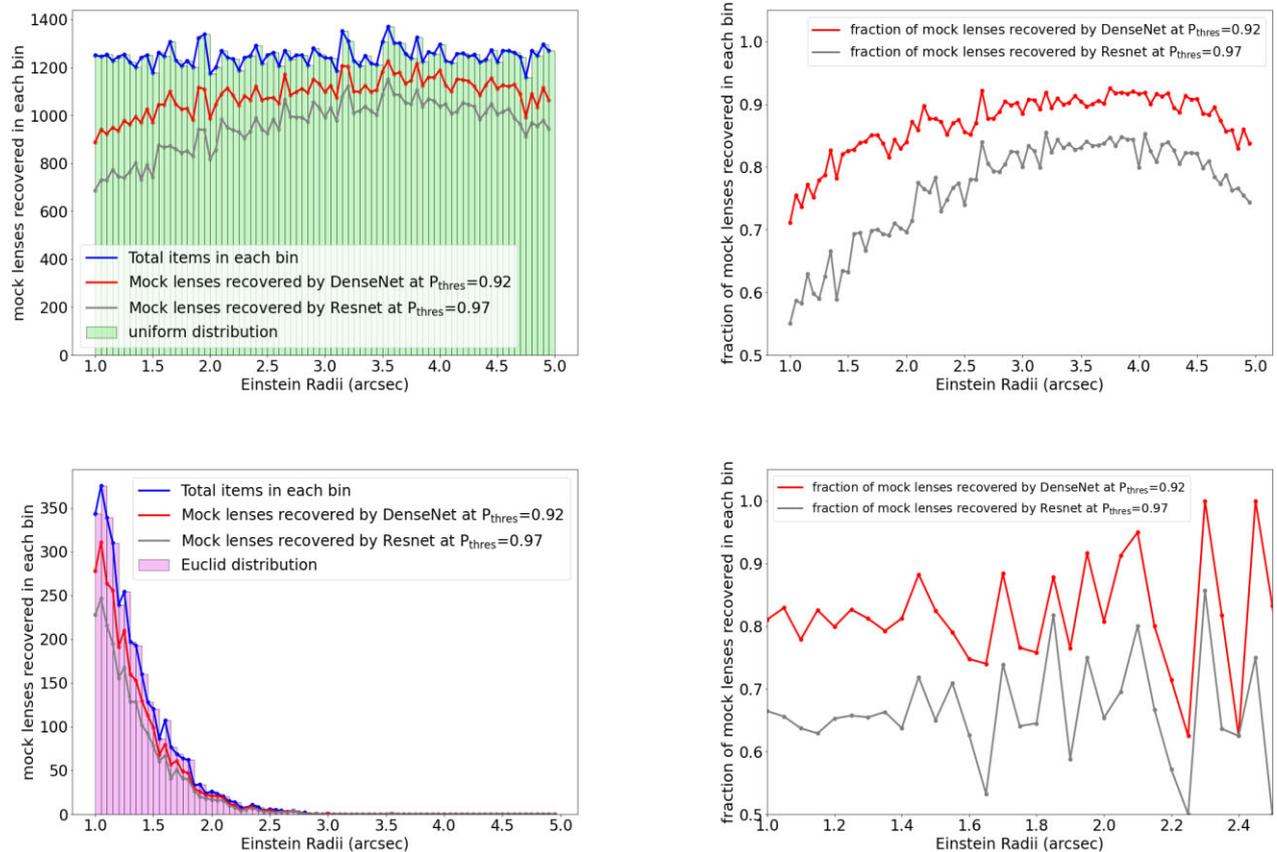

**Figure 7.** *Top left:* 100 000 mock lenses were generated with their Einstein radius uniformly distributed between the range [1,5] arcsec which is shown as a green histogram marked by the blue line. The mock lenses recovered by DenseNet and ResNet are shown in red and grey, respectively. *Top right:* The fraction of mock lenses recovered (out of the total number of lenses) by DenseNet and ResNet networks are shown in red and grey, respectively. *Bottom left:* 3379 mock lenses are sampled from uniform distribution showing Einstein radius distribution (purple histogram) of galaxy–galaxy strong lenses that will be discovered in *Euclid* (Collett 2015). The blue continuous line shows the total items present in each bin. Red and grey lines show the number of mock lenses which can be discovered by both DenseNet and ResNet ensembles at constant FPR as shown as points 'a' and 'b' in Fig. 4, i.e., DenseNet at $P_{thres} = 0.92$ and ResNet at $P_{thres} = 0.97$. *Bottom right:* The fraction of mock lenses (out of total mock lenses available in each bin) discovered by DenseNet and ResNet is shown in red and grey lines, respectively.

IC values help to select the most information-rich lens candidates among all the classified mock lenses.

### 4.4 DenseLens prediction for a more realistic Einstein radius distribution

To further investigate what we can expect from our 'DenseLens' network, we generate 100 000 mock lenses with their Einstein radius distributed uniformly in the range [1,5] arcsec, as shown in Fig. 7 (top left panel). The blue continuous line represents the number of mock lenses present in each bin. The number of mock lenses recovered by DenseNet and ResNet ensemble networks is shown in red and grey lines, respectively. Their respective fraction of mock lenses recovered is shown in the top right panel. We can observe that the DenseNet-121 ensemble recovers more mock-lenses when compared to the ResNet-18 ensemble irrespective of the Einstein radii, confirming our earlier conclusion but showing that this holds for lens systems of any Einstein radius. The overall TPR seems to be higher for both the models when compared to Fig. 4. This is due to the fact that the Einstein radii distribution is uniform (in Fig. 7; top left) and thus the overall TPR seems to be higher, whereas in Fig. 4, the test data set distribution is dominated by lenses in Einstein radii of 1–2 arcsec and thus the TPR is relatively low. It is interesting to observe that the fraction of mock lenses recovered is the highest in the interval of 3–4 arcsec. One plausible explanation is that at higher Einstein radii (greater than 4 arcsec), the size of the lensing features starts to closely resemble the features of negatives (such as spirals or galaxy groups), and hence the fraction of lenses recovered decreases in this interval.

In total, 3379 mock lenses (test data set) are sampled from this uniform distribution to mimic the Einstein radius distribution of galaxy–galaxy lenses as expected for the Euclid mission (Collett 2015), although this distribution is expected to be very similar for other surveys such as KiDS. The histogram of mock lenses having the Einstein radii distribution is shown in Fig. 7 (bottom left), with a lower limit of 1 arcsec. The blue continuous line shows the total number of mock lenses present in each bin. The red and grey continuous lines represent the mock lenses discovered at a constant FPR as labelled by points 'a' and 'b' in Fig. 4, i.e. by DenseNet at $P_{thres} = 0.92$ and ResNet, respectively, at $P_{thres} = 0.97$. Their respective fraction of mock lenses recovered is shown in the bottom right. In the interval of 1–2 arcsec, where most lenses are expected to be found, the DenseNet-121 ensemble network clearly recovers more mock lenses than the ResNet-18 ensemble network, in fact, this is the case for all Einstein radii as shown in Fig. 7 (top and bottom right panels). Since Euclid will have better resolution when compared to KiDS, we will train the DenseNet-121 ensemble network again with *Euclid*-like images again for analysing *Euclid* data.

## 5 SUMMARY AND CONCLUSIONS

We have introduced the use of a DenseNet architecture to find strong lenses and compared its performance to ResNet architecture and customized ResNet used in Li et al. (2021). We have trained four independent DenseNet and ResNet CNNs for our classification problem using KiDS *r*-band data of LRGs, combined with simulated lensed images as in Petrillo et al. (2017, 2019a, b) and Li et al. (2020).

Given the unbalanced nature of the real observational data set, the number of mock lenses found can be compared one to one with non-lenses at a FPR rate of $10^{-3}$. At this low FPR rate of $10^{-3}$, the highest TPR achieved among all networks is 0.68. We find in our study of comparing three networks at constant TPR equal to 0.68, that the DenseNet-121 ensemble can recover half the number of false positives when compared with the Li (2021) ResNet+ network and seven times fewer false positives when compared with the ResNet-18 ensemble network when trained with the same data and same number of training iterations. Similarly, at a constant, very low value of the FPR of $10^{-4}$, the DenseNet-121 ensemble can recover 3 per cent more mock lenses than Li (2021) ResNet+ and 6 per cent more mock lenses than ResNet-18 ensemble networks. Finding more mock lenses at very low FPR rates can be incredibly beneficial in upcoming large survey missions such as *Euclid*. More importantly, DenseNet-121 ensemble achieves this performance with ten times and three times fewer parameters when compared with ResNet-18 ensemble networks and Li (2021) ResNet+ network, respectively.

We have introduced the concept of rank-ordering classified images based on their IC in addition to rank-ordering them on *P*-values. We have defined IC as a value that scales linearly with the number of spatial resolution elements of the mock-lens above a given brightness threshold in units of the background noise and Einstein radius ($R_E$) over the smaller effective source radius ($R_{eff}$). We have shown that rank ordering of lens candidates can be done with the predicted IC values for test images, reducing visual inspection in the process. This will be particularly important in future surveys such as with *Euclid* where human inspection of the results is no longer feasible.

We have developed a pipeline ensemble model, called 'DenseLens', consisting of both classification and regression CNNs. We have shown that the Classification ensemble model of DenseLens can be used initially in the pipeline to filter out images whose classification prediction scores ($P_{mean}$) are less than the threshold value. The selected images are then passed through the regression ensemble network which predicts the IC values for the test images. Based on the IC images, we have rank-ordered the final candidates.

The actual IC values and the predicted IC values have a good correlation and the candidates having high estimated IC values and true IC values equal to zero are separated by the classification ensemble network. The majority of the candidates having estimated IC values greater than 100 are indeed correctly classified ($P_{mean} > 0.94$) by the classification ensemble network.

In addition to this, we have also generated test sets with a realistic distribution of Einstein radii, showing that the DenseNet-121 ensemble network recovers more mock lenses when compared to the ResNet-18 ensemble network for all Einstein radii.

In the future, the training data can be improved by making it more realistic by using more realistic parameter distributions in generating mock lenses. Also other lensing types such as quasar-like lensing features can be added to our set of mock lenses. Increasing the data base of our negatives such as spiral galaxies from the new KiDS release should also be done in order to even further reduce the false positive rate, which is still considerable for $P < 0.94$. In the future, we can also train individual CNNs for each lensing feature type as different classes instead of binary classification.

Finally, in a forthcoming paper, we will use the DenseLens Pipeline-Ensemble model to find new lenses from the KiDS data release four (DR4) and five (DR5).

## ACKNOWLEDGEMENTS

We would like to thank the Center for Information Technology of the University of Groningen for their support and for providing access to the Peregrine high performance computing cluster. The research for this paper was funded by the Centre for Data Science and Systems Complexity at the University of Groningen (www.rug.nl/research/fse/themes/dssc/).

## 6 DATA AVAILABILITY

The data used in the paper are available on request. The data underlying this article will be shared on reasonable request to the corresponding author.

## REFERENCES

Alexander S., Gleyzer S., McDonough E., Toomey M. W., Usai E., 2020, ApJ, 893, 15
Barnabè M., Czoske O., Koopmans L. V. E., Treu T., Bolton A. S., Gavazzi R., 2009, MNRAS, 399, 21
Barnacka A., 2018, Phys. Rep., 778, 1
Biesiada M., 2006, Phys. Rev. D, 73, 023006
Bleem L. E. et al., 2015, ApJS, 216, 27
Bolton A. S., Burles S., Koopmans L. V. E., Treu T., Moustakas L. A., 2006, ApJ, 638, 703
Bolton A. S., Burles S., Koopmans L. V. E., Treu T., Gavazzi R., Moustakas L. A., Wayth R., Schlegel D. J., 2008, ApJ, 682, 964
Browne I. et al., 2003, MNRAS, 341, 13
Capaccioli M., Schipani P., 2011, Messenger, 146, 27
Chan J. H. H. et al., 2016, ApJ, 832, 135
Collett T. E., 2015, ApJ, 811, 20
Congdon A. B., Keeton C. R., 2018, Strong Lensing by Galaxies. Springer International Publishing, Cham, p. 145
Davies A., Serjeant S., Bromley J. M., 2019, MNRAS, 487, 5263
de Jong J. T., Verdoes Kleijn G. A., Kuijken K. H., Valentijn E. A., 2013, Exp. Astron., 35, 25
Dewdney P. E., Hall P. J., Schilizzi R. T., Lazio T. J. L. W., 2009, Proc. IEEE, 97, 1482
Diehl H. T. et al., 2017, ApJS, 232, 15
Eisenstein D. J. et al., 2001, AJ, 122, 2267
Faure C. et al., 2008, ApJS, 176, 19
Gavazzi R., Treu T., Rhodes J. D., Koopmans L. V. E., Bolton A. S., Burles S., Massey R. J., Moustakas L. A., 2007, ApJ, 667, 176
Gentile F. et al., 2021, MNRAS, 510, 500
Glorot X., Bordes A., Bengio Y., 2011, In Gordon G., Dunson D., Dudík M., eds, Proceedings of the Fourteenth International Conference on Artificial Intelligence and Statistics, Vol. 15, Proceedings of Machine Learning Research. PMLR, Fort Lauderdale, p. 315
Goyal S., Harikrishnan. D., Kapadia S. J., Ajith P., 2021, Phys. Rev. D, 104, 124057
Grillo C. et al., 2018, ApJ, 860, 94
Halkola A., Seitz S., Pannella M., 2006, MNRAS, 372, 1425

He K., Zhang X., Ren S., Sun J., 2016, 2016 IEEE Conference on Computer Vision and Pattern Recognition (CVPR). Deep Residual Learning for Image Recognition, pp. 770–778.
Heymans C. et al., 2012, MNRAS, 427, 146
Huang G., Liu Z., van der Maaten L., Weinberger K. Q., 2017, 2017 IEEE Conference on Computer Vision and Pattern Recognition (CVPR). pp. 2261–2269.
Ioffe S., Szegedy C., 2015, Proceedings of the 32nd International Conference on Machine Learning. In Francis B., and Blei D.eds, Vol. 37, Lille, France. pp. 448–456.
Jackson N. J., 2008, MNRAS, 389, 1311
Jacobs C. et al., 2019a, ApJS, 243, 17
Jacobs C. et al., 2019b, MNRAS, 484, 5330
Jones C. M., Athanasiou T., 2005, Ann. Thoracic Surg., 79, 16
Kingma D. P., Ba J., 2017, Adam, A Method for Stochastic Optimization, preprint(arXiv:1412.6980)
Kochanek C. S., Schechter P. L., 2003, Carnegie Observatories Astrophysics Series, 2, 211
Kochanek C. S., 2006, Strong Gravitational Lensing. Springer, Berlin, p. 91
Koopmans L. V. E. et al., 2009, ApJ, 703, L51
Kormann R., Schneider P., Bartelmann M., 1994, A&A, 284, 285
Krizhevsky A., Sutskever I., Hinton G. E., 2012, In Pereira F., Burges C. J. C., Bottou L., Weinberger K. Q., eds, Advances in Neural Information Processing Systems. Vol. 25, Curran Associates, Inc. New York, NY, United States
Kuijken K., et al., 2011, Messenger, 146
Kuijken K. et al., 2019, A&A, 625, A2
Laureijs R. J., Duvet L., Sanz I. E., Gondoin P., Lumb D. H., Oosterbroek T., Criado G. S., 2010, In J. M. O., Jr, Clampin M. C., MacEwen H. A., eds, Proc. SPIE Conf. Ser. Vol. 7731, Space Telescopes and Instrumentation 2010: Optical, Infrared, and Millimeter Wave. SPIE, Bellingham, p. 453
Lecun Y., Bottou L., Bengio Y., Haffner P., 1998, Proc. IEEE, 86, 2278
Li R., Shu Y., Wang J., 2018, MNRAS, 480, 431
Li R. et al., 2020, ApJ, 899, 30
Li R. et al., 2021, ApJ, 923, 16
Linder E. V., 2004, Phys. Rev. D, 70, 043534
Lintott C. J. et al., 2008, MNRAS, 389, 1179
Lintott C. et al., 2011, MNRAS, 410, 166
McKean J. P. et al., 2007, MNRAS, 378, 109
Melvin T. et al., 2014, MNRAS, 438, 2882
Meneghetti M., Bartelmann M., Dolag K., Moscardini L., Perrotta F., Baccigalupi C., Tormen G., 2005, A&A, 442, 413
Metcalf R. B. et al., 2019, A&A, 625, A119
Miyazaki S. et al., 2012, In McLean I. S., Ramsay S. K., Takami H., eds, Proc. SPIE Conf. Ser. Vol. 8446, Ground-Based and Airborne Instrumentation for Astronomy IV. SPIE, Bellingham, p. 327
More A. et al., 2016, MNRAS, 465, 2411
Narayan S., 1997, Inf. Sci., 99, 69
Nightingale J. W., Massey R. J., Harvey D. R., Cooper A. P., Etherington A., Tam S.-I., Hayes R. G., 2019, MNRAS, 489, 2049
Nord B. et al., 2016, ApJ, 827, 51
Oguri M. et al., 2006, AJ, 132, 999
Oguri M. et al., 2008, AJ, 135, 520
Pawase R. S., Courbin F., Faure C., Kokotanekova R., Meylan G., 2014, MNRAS, 439, 3392
Pearson J., Pennock C., Robinson T., 2018, Emergent Sci., 2, 1
Petrillo C. E. et al., 2017, MNRAS, 472, 1129
Petrillo C. E. et al., 2019a, MNRAS, 482, 807
Petrillo C. E. et al., 2019b, MNRAS, 484, 3879
Quinn P., Axelrod T., Bird I., Dodson R., Szalay A., Wicenec A., 2015, Delivering SKA science, preprint(arXiv:1501.05367)
Rezaei S., McKean J., Biehl M., de W., Lafontaine A., 2022, A machine learning based approach to gravitational lens identification with the International LOFAR Telescope, preprint (arXiv:2207.10698)
Rhee G., 1991, Nature, 350, 211
Richard J. et al., 2014, MNRAS, 444, 268
Rojas K. et al., 2022, A&A, 668, A73
Rosen B. E., 1996, Connect. Sci., 8, 373
Sarbu N., Rusin D., Ma C.-P., 2001, ApJ, 561, L147
Sereno M., 2002, A&A, 393, 757
Serjeant S., 2014, ApJ, 793, L10
Sérsic J., 1968, Bull. Astron. Inst. Czech., 19, 105
Shu Y. et al., 2015, ApJ, 803, 71
Shu Y. et al., 2017, ApJ, 851, 48
Spiniello C., Barnabè M., Koopmans L. V. E., Trager S. C., 2015, MNRAS, 452, L21
Szegedy C. et al., 2014, Going deeper with convolutions, preprint (arXiv:1409.4842)
Tanaka M., 2016, ApJ, 826, L19
The Dark Energy Survey Collaboration. 2005. The Dark Energy Survey, preprint (astro-ph/0510346)
Treu T., 2010, ARA&A, 48, 87
Treu T., Koopmans L. V. E., 2002, ApJ, 575, 87
Treu T., Koopmans L. V. E., 2004, ApJ, 611, 739
Treu T. et al., 2018, MNRAS, 481, 1041
Tyson J. A., 2002, in Tyson J. A., Wolff S., eds, Proc. SPIE Conf. Ser. Vol. 4836, Survey and Other Telescope Technologies and Discoveries. SPIE, Bellingham, p. 10
Verdugo T., de Diego J. A., Limousin M., 2007, ApJ, 664, 702
Willett K. W. et al., 2013, MNRAS, 435, 2835
Zhan H., 2018, 42nd COSPAR Scientific Assembly, 42, E1
Zitrin A. et al., 2012, ApJ, 749, 97

## APPENDIX A: CREATION OF LENS EXAMPLES

Lens examples are created by combining an LRG with a simulated mock lensed source as shown in the Fig. A1

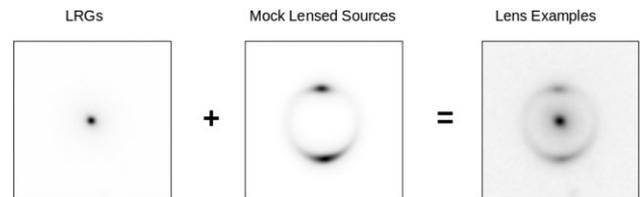

**Figure A1.** Illustration of lens examples creation. Lens examples are created by combining an LRG from KiDS data set with a mock lensed source.

## APPENDIX B: DENSELY CONNECTED CONVOLUTIONAL NETWORKS

DenseNets (Huang et al. 2017) uses an architecture in which each layer is connected with the next layer in a feed-forward manner. Each layer uses feature maps of all previous layers as inputs. This type of architecture results in various advantages such as encouraging feature reuse, thereby strengthening feature propagation and resulting in a lesser number of parameters. DenseNet also has been proven to have better parameter utilization when compared with ResNets and thus they are less prone to overfitting of the model. For 'n' layers in the DenseNet architecture, $n^{th}$ layer receives feature maps of all preceding layers as its input. We have used [6, 12, 24, 16] layers for the dense block as shown in the paper for the DenseNet-121 architecture. We have set the growth rate of network (k) to 12 and we have used bottleneck layers and compression as shown in the paper to improve computational efficiency and to reduce model size. Further, we have used bottleneck layers and compression as mentioned in the paper for DenseNet-BC architecture. Hence, in short, we have used a DenseNet-BC architecture with 121 layers with a growth rate (k) equals 12.

$$x_n = H_n([x_0, x_1 ... x_{n-1}]) \tag{B1}$$

where $H_n(.)$ is an composite function comprising of batch normalization layer (Ioffe & Szegedy 2015) followed by a Rectified Linear Unit (ReLU) (Glorot, Bordes & Bengio 2011) activation layer and $3 \times 3$ convolution layer.

## APPENDIX C: RESIDUAL NEURAL NETWORKS

ResNets (He et al. 2016) address the problem of difficulty in training deeper neural networks and hence come with a solution of residual learning framework. In the residual learning framework, identity mapping is performed by shortcut connection as shown in the Fig. C1. We use the paper 18 layer ResNet architecture (ResNet18) for comparing ResNet performance with DenseNet. ResNets were widely used in classifying strong lenses by Petrillo et al. (2017, 2019a, b) and Li et al. (2020, 2021).

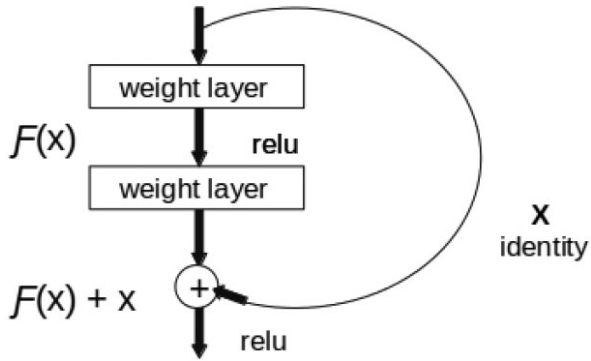

**Figure C1.** Illustration of residual learning block in ResNets (He et al. 2016). The above figure illustrates on how each layer is connected to all subsequent layers.

## APPENDIX D: TRAINING THE CNN

For this paper, we have used comparison of DenseNet and ResNet type of architectures. For both the networks, the training was done for 2500 gradient update steps and at each step 512 images (256 lenses, 256 non-lenses) were used for training. Thus approximately 1.28 M images were used to train the network. Also at each step, 256 images were used for validation. The training (top) and validation (bottom) step versus loss for single classification DenseNet-121 and ResNet-18 network is shown in Fig. D1. The DenseNet-121 architecture achieves lowest loss in a few hundred training steps which ResNet-18 architecture fails to achieve even after 2500 iterations. This is only attributed due to the difference in network architecture types between ResNet-18 and DenseNet-121. The training of each DenseNet-121 classification network for 2500 steps took ∼13.7 h and used an average ∼14.5 per cent of RAM allocated on 120 GB allocated 'NVIDIA V100' GPU (Graphics Processing Unit). Whereas, ResNet-18 type architecture took only

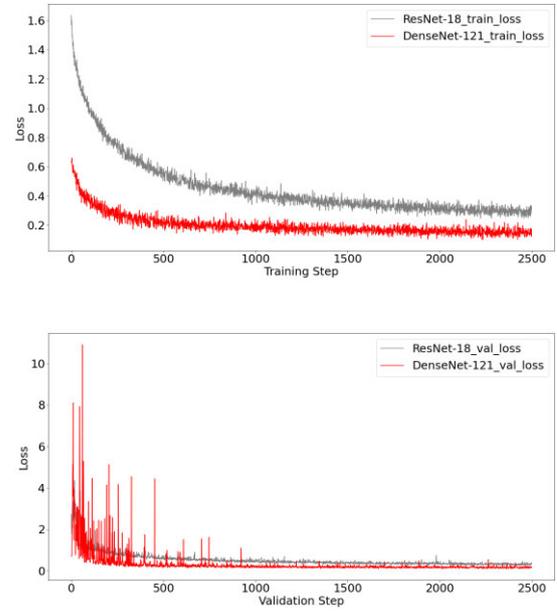

**Figure D1.** *Top*: Training loss versus step for a single DenseNet-121 and ResNet-18 network. *Bottom*: Validation loss versus step for a single DenseNet-121 and ResNet-18 network. DenseNet-121 achieves training loss quicker when compared to ResNet-18 network.

∼2.85 h for training of 2500 steps and ∼13.15 per cent of RAM was used on average out of 120 GB allocated on the same GPU machine. We have to note that, even though DenseNet-121 architecture type uses more memory and takes more time to train for the same number of iterations when compared to ResNet-18 architecture, it achieves the lower training error within a few hundred steps, which ResNet-18 architecture fails to achieve even after training for 2500 steps. For *Classification Networks* (ensemble of CNNs 1–4), we minimize the Binary Cross Entropy (BCE) loss function using ADAM optimizer (Kingma & Ba 2017) with a learning rate of 0.001. For *Regression Networks* (ensemble of CNNs 5–8), we minimize mean absolute error loss with the same ADAM optimizer and the same learning rate of 0.001.

## APPENDIX E: ROC CURVES OF INDIVIDUAL CNNS

The ROC curves of ensemble of four DenseNet-121 and ResNet-18 architectures are shown in Fig. 4. In Fig. E1, we have shown the individual ROC curves of the DenseNet-121, ResNet-18 CNNs and Li (2021) ResNet+ with the same test data used to generate the ROC curve shown in Fig. 4. We clearly see that the Area Under ROC (AUROC) for each individual DenseNet-121 architecture is higher when compared to the ResNet-18 architectures. ROC curve of Li (2021) ResNet+ outperforms all ResNet-18 individual networks and some DenseNet-121 individual networks because of the parameter heavy additional dense layers in the end. In general, higher the values of the AUROC imply better performance of the model. Thus, we can argue that the DenseNet-121 architecture outperforms or produces similar results to ResNet-18 network architecture with 10 times fewer parameters.

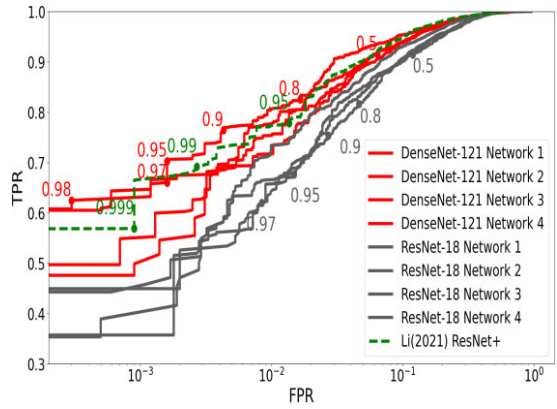

**Figure E1.** ROC curves of individual DenseNet-121 and ResNet-18 CNNs.

# APPENDIX F: OTHER APPROACHES

We have examined different approaches with CNNs for finding strong lenses. We found that the following approaches were performing poorer when compared to the Pipeline-Ensemble Model shown in Section 3.2.

# APPENDIX G: HISTOGRAM OF P VALUES

The histogram of classification prediction scores of DenseNet Networks 1–4 and its ensemble is shown in Fig. G1 for mock lenses (top) and non-lenses (bottom). For mock lens samples (shown at the top), the correlation between DenseNet Network 1–4 is less for lower values of P and similarly, for non-lens samples (shown at the bottom), the correlation among DenseNet Network 1–4 is less for higher values of P.

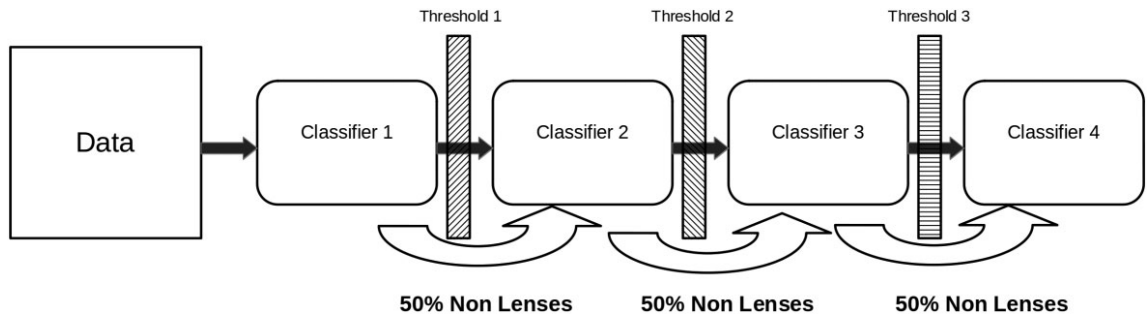

**Figure F1.** Illustration of cascade classifier.

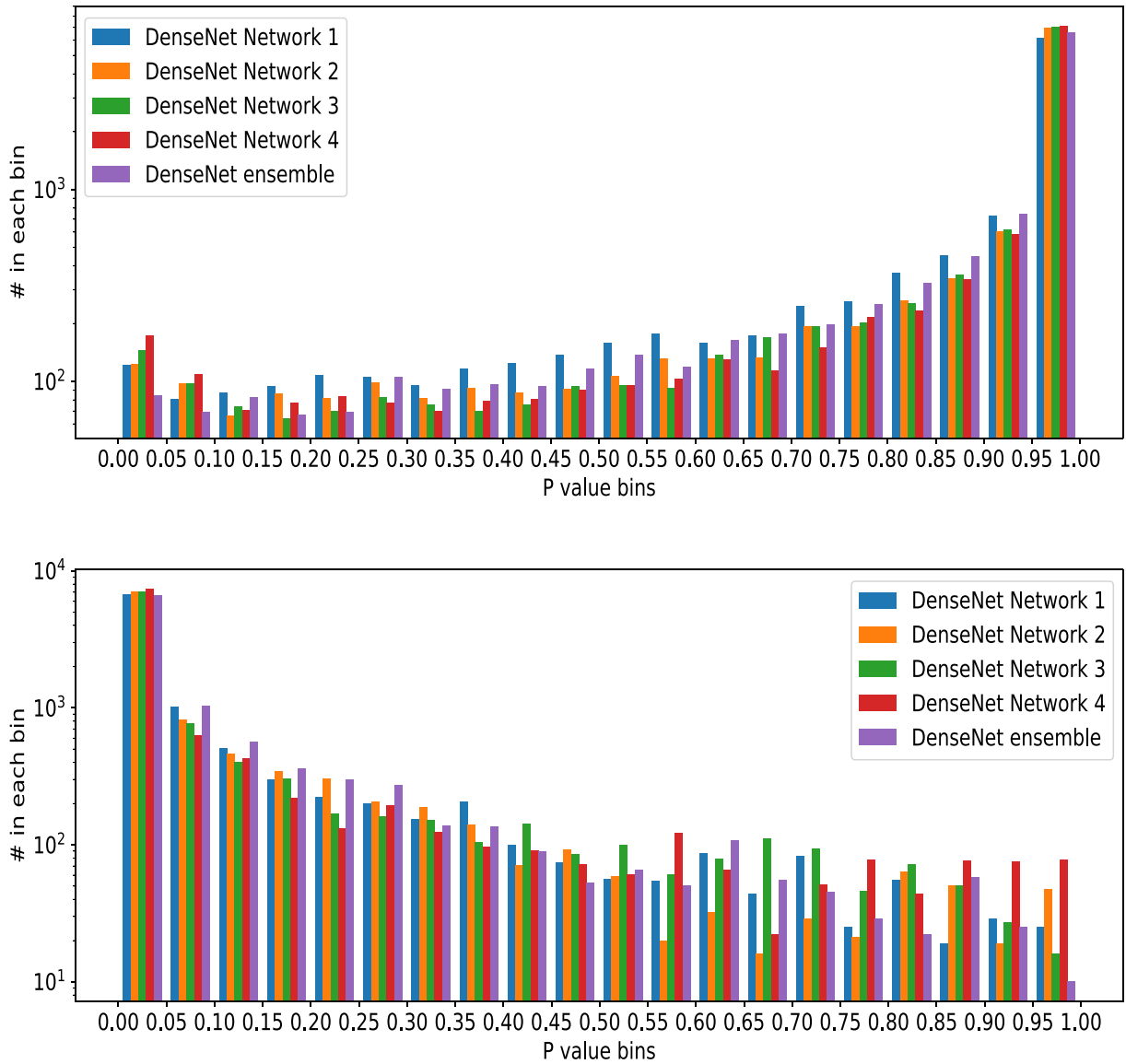

**Figure G1.** The histogram of classification prediction scores (P) of DenseNet Networks 1–4 and its ensemble for mock lens samples (top) and non-lens samples (bottom).

This paper has been typeset from a T<sub>E</sub>X/L<sup>A</sup>T<sub>E</sub>X file prepared by the author.